# Penalized Linear Models for Highly Correlated High-Dimensional Immunophenotyping Data

**Authors:** Xiaoru Dong[1], Apoorva Goyal[1,2], Muxuan Liang[1], Maigan A. Brusko[2,3], Todd M. Brusko[2,3], Rhonda Bacher[1,2] *

**Affiliations:**

[1]Department of Biostatistics, College of Public Health and Health Professions, University of Florida, Gainesville, FL 32610, USA.

[2]Diabetes Institute, University of Florida, Gainesville, FL 32610, USA.

[3]Department of Pathology, Immunology, and Laboratory Medicine, College of Medicine, University of Florida, Gainesville, FL 32610, USA.

*Corresponding author:

Rhonda Bacher, PhD, 2004 Mowry Rd, P.O. Box 117450 Gainesville, FL 326110. E-mail: rbacher@ufl.edu. TEL: (352) 294-5914. ORCID: 0000-0001-5787-476X

**Abstract**

Accurate prediction and identification of variables associated with outcomes or disease states are critical for advancing diagnosis, prognosis, and precision medicine in biomedical research. Regularized regression techniques, such as lasso, are widely employed to enhance interpretability by reducing model complexity and identifying significant variables. However, when applying to biomedical datasets, e.g., immunophenotyping dataset, there are two major challenges that may lead to unsatisfactory results using these methods: 1) high correlation between predictors, which leads to the exclusion of important variables with included predictors in variable selection, and 2) the presence of skewness, which violates key statistical assumptions of these methods. Current approaches that fail to address these issues simultaneously may lead to biased interpretations and unreliable coefficient estimates. To overcome these limitations, we propose a novel two-step approach, the Bootstrap-Enhanced Regularization Method (BERM). BERM outperforms existing two-step approaches and demonstrates consistent performance in terms of variable selection and estimation accuracy across simulated sparsity scenarios. We further demonstrate the effectiveness of BERM by applying it to a human immunophenotyping dataset identifying important immune parameters associated the autoimmune disease, type 1 diabetes.

**Introduction**

With biomedical data increasingly occupying high-dimensional spaces, the potential for precision medicine and identifying predictive biomarkers is more achievable than ever before. Techniques such as mass cytometry, high-dimensional flow cytometry, and CITE-seq enable the characterization of cellular subsets and activation states based on 20 to 50 individual surface and intracellular markers measured simultaneously (Delmonte & Fleisher, 2019; Thomas et al., 2017; Yao et al., 2022). The combination or interaction of these parameters, representing hundreds to thousands of cellular variables, is of particular interest in identifying disease-associated cellular signatures. Immune cells, for instance, are broadly defined in peripheral blood by canonical cell surface markers, including CD3 (T cells), CD19 (B cells), CD11c (dendritic cells [DCs]), CD56 (natural killer [NK] cells), and CD68 (macrophages). Combinations of these and other markers further distinguish different cell subsets and functional statuses (Dutertre et al., 2019), enabling the identification of alterations in cell compositions associated with vaccination, infection, autoimmunity, and cancer (Bergman et al., 2022; Golovkin et al., 2021; Khunger et al., 2021; Perry et al., 2020). As the number of parameters increases with technological advances in instrumentation and reagents, the analysis of immunophenotyping data increasingly requires advanced statistical methods. For instance, the high dimensionality leads to challenges in identifying and quantifying which systematic variations in cell states are associated with time, lifespan, and disease (Shapiro et al., 2023). With immunophenotyping data in particular, the primary goal is to uncover underlying patterns that inform cellular function and predict disease progression, as well as enhance our understanding of immune system dynamics. In this context, an analysis technique that focuses on both variable selection and coefficient estimation to quantify the effects of variables is preferred.

For high-dimensional settings, regression techniques, such as the lasso (Tibshirani, 1996) and adaptive lasso (Zou, 2006), have been widely utilized to select important variables and estimate their effects on an outcome by penalizing individual coefficients. However, the dependent nature of cellular parameters introduces a high degree of correlation among variables.

Lasso has previously been shown to inadequately handle highly correlated parameters, often selecting only one predictor among the correlated ones. To address this, the elastic net (Zou & Hastie, 2005) and adaptive elastic net (Zou & Zhang, 2009) were developed to leverage the best properties of the lasso and ridge penalties, resulting in the ability to select multiple correlated variables in high-dimensional settings. Additional methods have since been developed to tackle more complex correlation scenarios (e.g., varying magnitude or direction) such as random lasso (Wang et al., 2011), Hi-LASSO (Y. Kim et al., 2019), the nonparametric bootstrap quantile method (QNT) (Abram et al., 2016), and the Robust Elastic Net via Bootstrap (RENBOOT) (H. Kim & Lee, 2021). Importantly, however, the latter two approaches do not directly address coefficient estimation, focusing instead on the selection and identification of important variables.

Despite substantial progress, accurate coefficient estimation following variable selection remains a challenging issue in complex multi-collinearity scenarios (Gregorich et al., 2021). Immunophenotyping datasets are not only characterized by high multi-collinearity and dependence, but also parameters tend to have substantially skewed distributions, which can further bias model estimates (Figure 1A). To address these challenges, we propose a new approach called the Bootstrap-Enhanced Regularization Method (BERM). BERM integrates bootstrapped confidence intervals with penalized regression techniques to obtain robust variable selection and coefficient estimation in highly correlated and complex multi-collinearity scenarios. In Section 2, we present our methodology and describe how it differs from existing approaches. Then, in Section 3, we demonstrate BERM's performance in simulation studies, where data are simulated with varying degrees of skewness, correlation, and sparsity. In Section 4, we apply our method to an immunophenotyping dataset comprised of 192 flow-cytometry-derived phenotypes on peripheral blood mononuclear cells. A discussion of our findings is presented in Section 5.

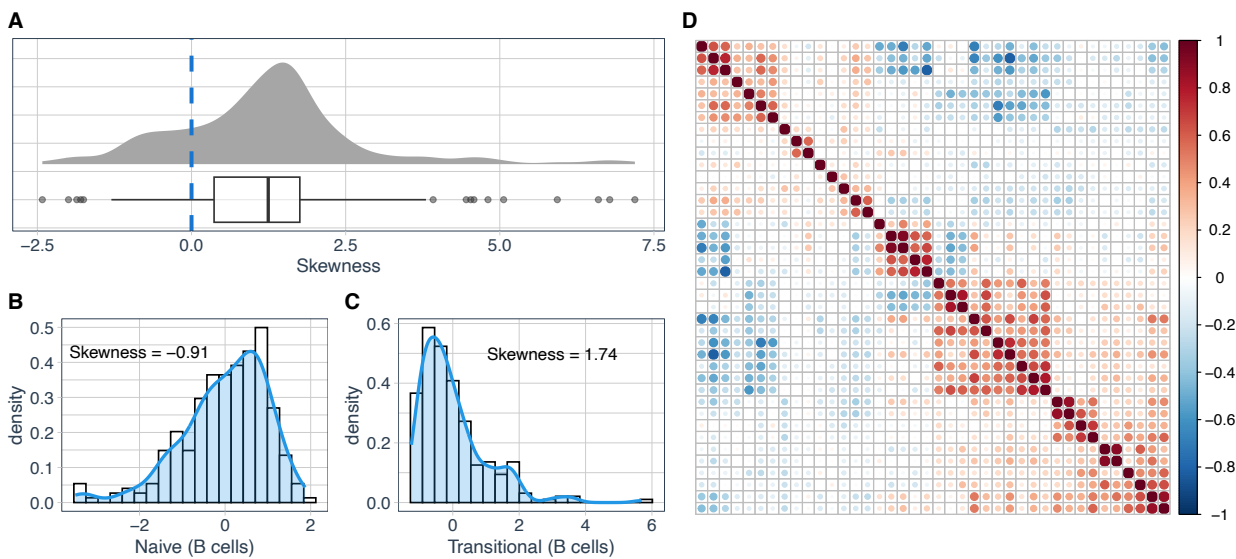

Figure 1. Skewness and correlation in published flow cytometric immunophenotyping data (Shapiro et al., 2023). A) Distribution of skewness for 192 immunophenotyping variables. Positive skewness indicates a right-skewed distribution, while negative skewness indicates a left-skewed distribution. B) Representation of negative skewness, exemplified by naïve B cells. C) Representation of positive skewness, exemplified by transitional B cells. D) Correlation matrix for a subset of significantly correlated variables from the immunophenotyping data. Positive correlations are displayed in various shades of red, and negative correlations are indicated by shades of blue.

## 2. Methods

### 2. 1 Problem Setup and Notation

Consider a human immunophenotyping dataset comprising $n$ individuals, each characterized by a response variable $y_i$ representing immune age, and the corresponding $p$-dimensional predictor vector $x_i = (x_{i1}, x_{i2}, \ldots, x_{ip})$ indicating the immunophenotyping variables. The relationship between the predictors and the response is generally modeled using the linear equation:

$$y_i = \beta_1 x_{i1} + \beta_2 x_{i2} + \ldots + \beta_p x_{ip} + \epsilon_i \qquad (1)$$

, where $\epsilon_i$ denotes the error term, assumed to be independently and identically distributed with $\epsilon|x$ having mean zero. The immunophenotyping variables are standardized, and the immune age is mean-corrected, eliminating the need for an intercept in the equation. Our goal is to identify a significantly associated subset of $p$ predictors and to obtain accurate coefficient estimates.

## 2. 2 BERM

We developed a two-step method which we call Bootstrap-Enhanced Regularization Method (BERM), that integrates techniques from both the adaptive elastic net and bootstrap confidence intervals. In the first step of BERM, an elastic-net based bootstrap confidence interval is obtained via data resampling. Variables with coefficients within a 95% confidence interval $CI_j = \left[\hat{\beta}_j^{0.025}, \hat{\beta}_j^{0.975}\right]$ that covers zero are deemed irrelevant. The relevance of a variable $x_j$ to the response variable is indicated by

$$r_j = \begin{cases} 1, & 0 \notin CI_j \\ 0, & \text{otherwise} \end{cases} \quad (2)$$

The second step of BERM incorporates the relevance of variables into a weighted elastic net model. The weighted penalty is included to achieve more accurate coefficient estimation by minimizing:

$$\min_\beta \frac{1}{2n} \sum_{i=1}^n \left(y_i - \sum_{j=1}^p \beta_j x_{ij}\right)^2 + w_j \lambda \left(\frac{1-\alpha}{2} \sum_{j=1}^p \beta_j^2 + \alpha \sum_{j=1}^p |\beta_j|\right) \quad (3)$$

, where

$$w_j = \begin{cases} 1, & r_j = 1 \\ \infty, & r_j = 0 \end{cases} \quad (4)$$

This approach helps to refine the accuracy of the coefficient estimates, thereby increasing the model's robustness and reliability, similar to the adaptive lasso (Zou, 2006). BERM is available as an R package on GitHub (https://github.com/xiaorudong/berm).

## 2.3 Fitting BERM

To optimize the computational efficiency of BERM, we set $\alpha = 0.5$, to evenly distribute the influence of the L1 and L2 penalties. Bootstrapping utilizes 100 resampling iterations by default. The train() function in the caret R package is used to tune the parameter $\lambda$ (Kuhn, 2008). Using the bootstrap samples, a 95% confidence interval with a 2.5th percentile lower bound and 97.5th percentile upper bound for coefficients is constructed. In the second step, the penalty arguments in the train() function are set to reflect the variable relevance determined in the first step.

**2.4 Differences from existing approaches**

Existing methods offer some strategies for regularization to optimize variable selection and coefficient estimation in the face of various modeling challenges. Both adaptive elastic net (Zou & Zhang, 2009) and adaptive lasso (Zou, 2006) incorporate weights into the penalty terms. However, the weights are based on the initial coefficient estimates instead of utilizing bootstrapping as in BERM. Other methods have addressed the challenge of high multi-collinearity by bootstrapping subsets of predictors in two stages, such as random lasso (Wang et al., 2011) and Hi-LASSO (Y. Kim et al., 2019), although both incur additional parameter-tuning burdens. The use of bootstrap intervals to determine variable importance was implemented by RENBOOT (H. Kim & Lee, 2021); however, coefficient estimation following variable selection was not addressed. In contrast, BERM fits an elastic net model with bootstrap-adapted weights for all predictors to systematically estimate the coefficients.

**3. Simulation Study**

**3. 1 Set-up**

Considering the dimensionality of realistic immunophenotyping datasets, we simulated data for two dimensionality settings: moderate-dimensional (large $n >$ moderate $p$) and high-dimensional (large $n \ll$ large $p$) spaces. For each setting, the true model was simulated according to three patterns of sparsity (i.e., the true proportion of zero coefficient variables): balanced (50% sparsity), a predominance of zero coefficients (75% sparsity), and a predominance of non-zero coefficients (25% sparsity). Additionally, simulated datasets were generated with varying noise levels, with $\sigma \in \{1, 3, 5\}$.

For the moderate-dimensional setting, the datasets consisted of 60 covariates and 300 observations. Non-zero coefficients were generated from $\beta \sim N(0, 4^2)$. To simulate datasets with multi-collinearity and skewness, the covariate matrix was generated as a multivariate non-normal distribution using the mnonr R package (Qu et al., 2020). Specifically, the multivariate skewness was set to 5,000, and multivariate kurtosis to 25,000. Multivariate skewness indicates deviation from the symmetry expected under a multivariate normal distribution, and multivariate kurtosis contributes to the heaviness of the distribution tails in multivariate space. The covariance matrix for each dataset was set as:

$$\begin{bmatrix} \Sigma^{20}_{U(0.6,1)} & 0 & 0 \\ 0 & \Sigma^{20}_{U(0.3,0.5)} & J^{20}_{0.2} \\ 0 & J^{20}_{0.2} & J^{20}_{0.05} \end{bmatrix}$$

, where $\Sigma^{k}_{U(a,b)}$ is a k × k symmetric matrix with unit diagonal elements and off-diagonal elements drawn from a uniform distribution U(a, b). $J^{k}_{v}$ is a k × k matrix with all elements value v. For generating covariance matrices indicative of different correlation strengths, $\Sigma^{20}_{U(0.6,1)}$ was used to reflect high correlation with elements from U(0.6, 1), while $\Sigma^{20}_{U(0.3,0.5)}$ corresponded to moderate correlation with elements from U(0.3, 0.5). Additionally, $J^{20}_{0.05}$ and $J^{20}_{0.2}$ were applied to introduce minimal correlation between variables.

For the high-dimensional setting, the datasets were simulated with 500 covariates and 300 observations. The covariate matrix was similarly generated from a multivariate non-normal distribution, but with a skewness of 10,000 and a kurtosis of 300,000 to ensure deviation from normality. The high-dimensional covariance matrix was generated similarly to the moderate-dimensional case with:

$$\begin{bmatrix} \Sigma_{U(0.6,1)}^{100} & 0 & 0 & 0 \\ 0 & \Sigma_{U(0.3,0.5)}^{100} & J_{0.1}^{100} & 0 \\ 0 & J_{0.1}^{100} & J_{0.1}^{100} & 0 \\ 0 & 0 & 0 & I^{200} \end{bmatrix}$$

, where $I^k$ is a k × k matrix with unit diagonal elements and off-diagonal element set to 0. Examples of data generated from these scenarios are illustrated Supplementary Figure 1.

Simulation of simple datasets

To assess the performance of each method affected by skewness and correlation, we generated "simple" datasets lacking both properties, using the same simulation settings described above. These datasets were generated with the rnorm() function, using a mean of zero and a standard deviation of one. Examples of data generated from these scenarios are illustrated in Supplementary Figure 2.

Simulation evaluation:

Balanced accuracy was used to evaluate the performance of each method on the simulated datasets in terms of variable selection accuracy.

$$\text{Balanced Accuracy} = 0.5 \times \left( \frac{TP}{TP + FN} + \frac{TN}{TN + FP} \right) \quad (5)$$

, where TP (True Positive) represents the number of correctly selected nonzero variables, TN (True Negative) represents the number of correctly non-selected zero variables, FP (False Positive) represents the number of falsely selected zero variables, and FN (False Negative) represents the number of non-selected nonzero variables.

The mean squared error (MSE) was used to assess each method's performance in terms of coefficient estimation. MSE was calculated only on the accurately selected variables for each method, and is defined as:

$$\text{MSE} = \frac{1}{p}\sum_{j=1}^{p}(\beta_j - \widehat{\beta_j})^2 \qquad (6)$$

, where $\beta_j$ represents the true coefficient value and $\widehat{\beta_j}$ denotes the estimated coefficient value for variable $j$.

**3.2 Simulation Results**

To understand how BERM and existing methods performed in identifying variables and estimating their effects, we simulated data matching the skewness and multi-collinearity characteristics of immunophenotyping data (Supplementary Figure 1). We also generated data across scenarios varying in dimensionality, noise level, and sparsity to represent a wide range of possibilities in biomedical datasets: moderate-dimensional settings (e.g., 60 variables and 300 observations) and high-dimensional settings (e.g., 500 variables and 300 observations). To account for a wide array of true models, we simulated three patterns of sparsity: balanced (50% sparsity), a predominance of zero coefficients (75% sparsity), and a predominance of non-zero coefficients (25% sparsity).

We compared BERM to traditional penalized models (Lasso, Elastic Net), their extensions (adaptive lasso, adaptive elastic net, RENBOOT), and methods specific to high multi-collinearity (random lasso and Hi-LASSO) (Details on set-up are provided in Supplementary Methods). Performance on feature selection accuracy was evaluated using balanced accuracy to account for the imbalance of zero and non-zero features. The evaluation of estimation bias was computed only on the accurately predicted variables for each method, to distinguish estimation from variable selection accuracy.

Accuracy in feature selection across simulation scenarios

We first compared the degree to which various methods either over- or under-selected features compared to the true number of important variables (sparsity-level). Methods largely differed in their performance across sparsity levels, although all methods tended to select more variables as the model sparsity increased (Figure 2). In moderate-dimensional settings, the adaptive methods and BERM performed the most consistently in selecting the accurate number of features across varying noise and sparsity levels (Figure 2A). Lasso, random lasso, and elastic net consistently over-selected features compared to the true sparsity level, while RENBOOT and Hi-LASSO under-selected features, particularly in low-sparsity scenarios. In the high-dimensional setting, most methods were prone to under-selection in the low- and mid-sparsity scenarios (Figure 2B). While most methods showed comparable performance across the moderate- and high-dimensional settings, standard elastic net and lasso performed remarkably better in the high-dimensional setting. In contrast, adaptive lasso performed worse in high dimensions, being overly conservative and consistently underestimating the number of relevant features.

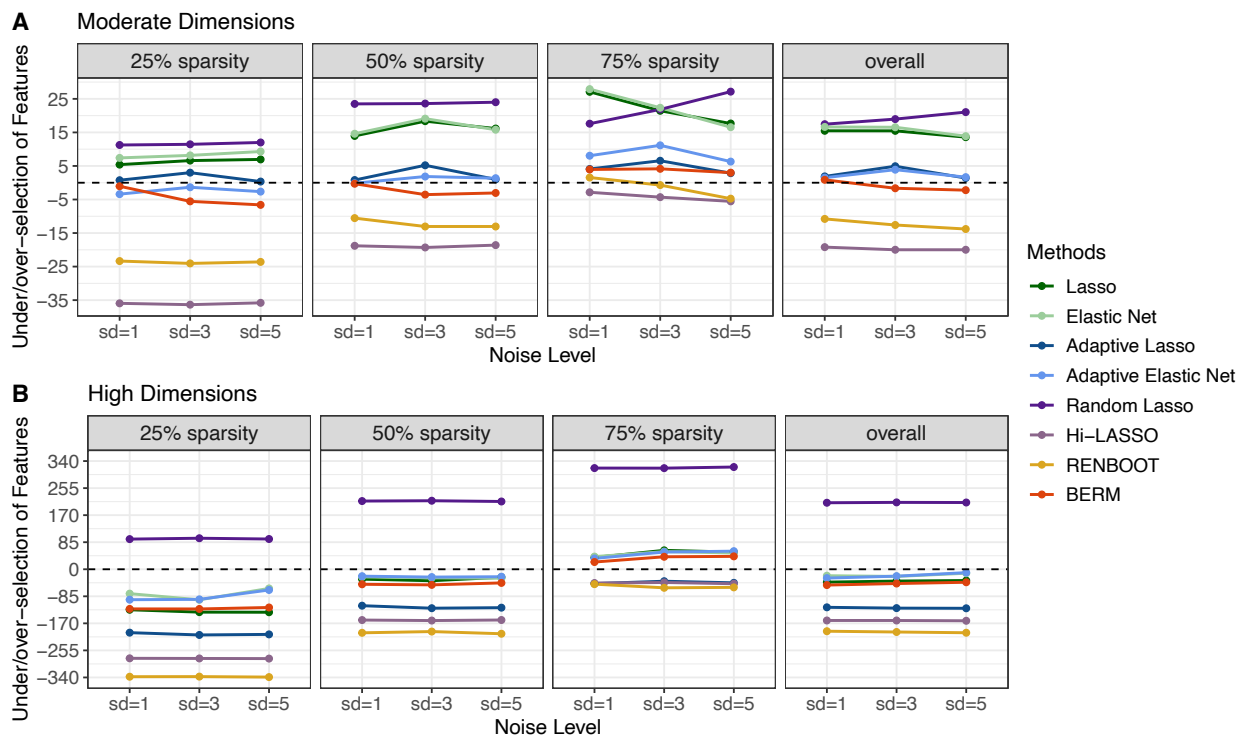

Figure 2. Assessment of under- and over-selection across moderate- and high-dimensional simulation scenarios. **A.** The average difference between the number of features selected and the true number of non-zero variables (y-axis) for moderate-dimensional data simulated under various noise levels (x-axis) and model sparsity levels (panels) is shown for eight different methods. **B.** Same as A, but for the high-dimensional scenario.

Next, we examined each method's variable selection accuracy in identifying truly non-zero features across the simulated scenarios (Figure 3). Relative to the average balanced accuracy across all scenarios, elastic net, lasso, and random lasso underperformed (ranging from -7.4% to -3.7% decrease in accuracy), the second-generation penalized methods (e.g. adaptive elastic net, adaptive lasso, RENBOOT and Hi-LASSO) performed slightly above average (ranging from +1.3% to +2.3% increase in accuracy), while BERM performed best (+10.6% accuracy). BERM had the largest average difference in accuracy between dimensionalities; outperforming all other methods in both settings. In particular, in the moderate-dimensional setting, BERM achieved an average of +17.0% higher accuracy relative to the overall mean performance (Figure 3A), with the next best performer, adaptive elastic net, showing a +5.2% increase. With the exception of RENBOOT, all methods had decreased performance as noise levels increased. Most declines in accuracy were modest (average -2.6% accuracy from low to high noise), with adaptive elastic net and adaptive lasso displaying the least stable performances across noise levels (difference of -5.5% and -4.5%, respectively). Most methods were stable across different sparsity levels, with accuracy differences ranging from -2.7% to +2.1%. However, the three lasso-based methods all performed substantially better in highly sparse scenarios compared to low-sparse scenarios (average of +6.5% to +13.3% increase in accuracy).

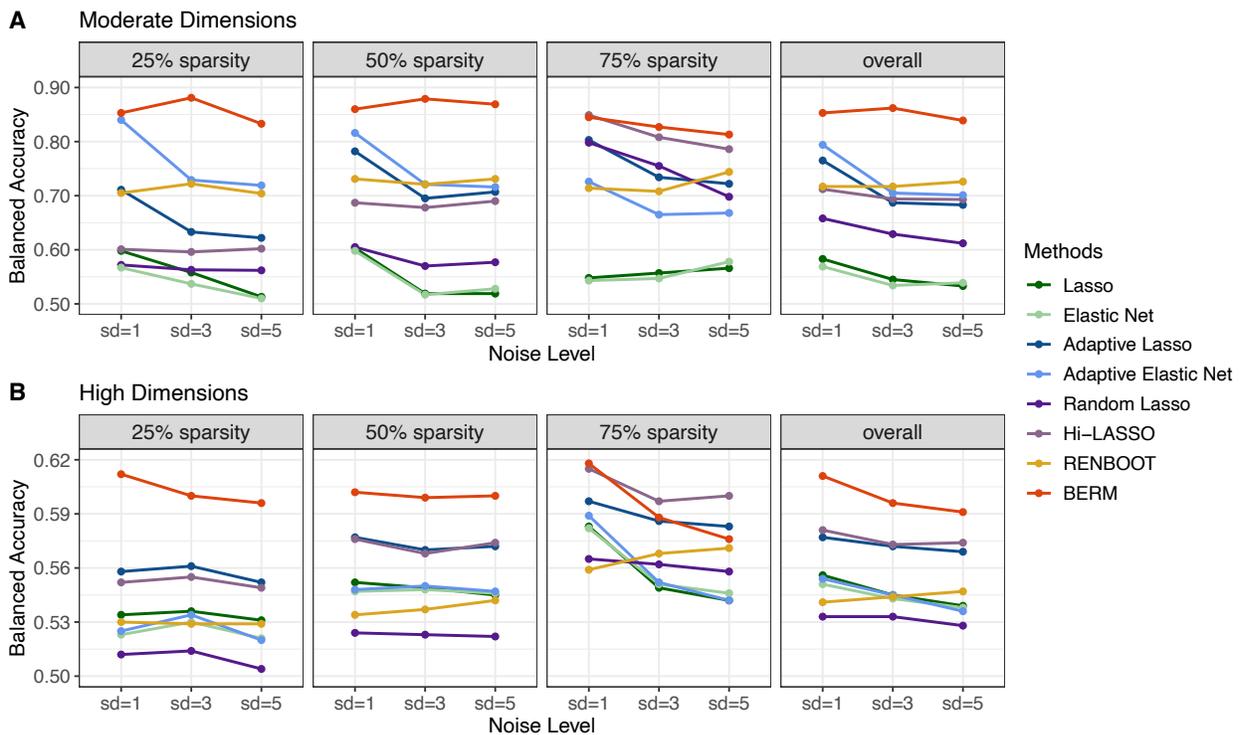

Figure 3. Feature selection accuracy across moderate- and high-dimensional simulated datasets. **A.** Mean balanced accuracy for correctly selecting non-zero variables (y-axis) in moderate-dimensional data, simulated under varying noise levels (x-axes) and model sparsity levels (panels), is shown for eight different methods. **B.** Same as A but for the high dimensional scenario.

Accuracy in coefficient estimation across simulation scenarios

The bias between the estimated coefficients and the true simulated coefficient values was calculated only on the accurately selected variables for each method (Figure 4). Except for random lasso and Hi-LASSO, all other methods exhibited similar overall levels of estimation error, which were largely unaffected by sparsity and noise levels across dimensionality settings. Adaptive lasso showed the largest discrepancy, with an average MSE of 8.0 in the high-dimensional setting compared to an MSE of 2.8 in the moderate-dimensional setting. Interestingly, random lasso and Hi-Lasso were also the only two methods to perform better in the high-dimensional setting compared to the moderate-dimensional, although they had the highest MSE in both settings.

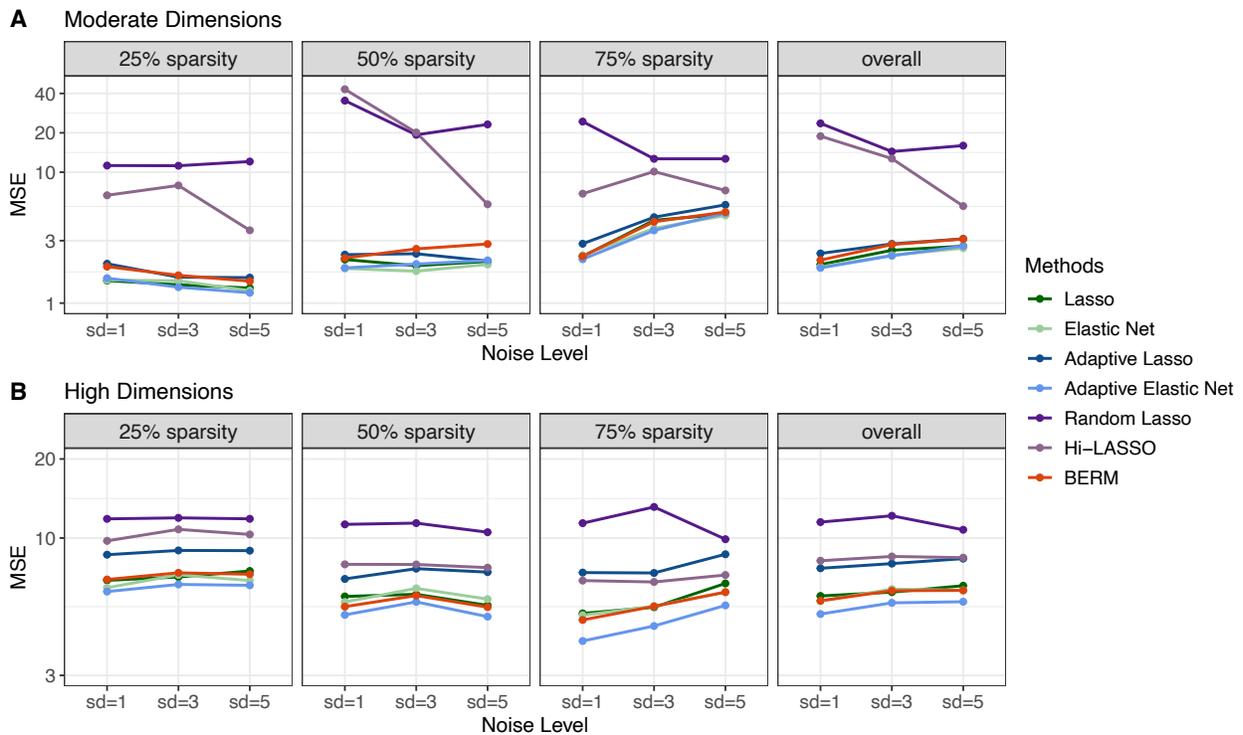

Figure 4. Bias in coefficient estimation across moderate- and high-dimensional simulated datasets. **A.** Mean squared error (MSE) calculated on accurately selected non-zero coefficients (y-axis) for moderate-dimensional data simulated under varying noise levels (x-axis) and model sparsity levels (panels) is shown for eight different methods. **B.** Same as A, but for the high-dimensional scenario.

Influence of skewness and correlation on method performance in feature selection and coefficient estimation

We next investigated how the specific characteristics of immunophenotyping data contributed to each method's performance. To do this, we generated "simple" simulated datasets that mirrored the complex simulated data but lacked skewness and high correlation among variables. Simulations were conducted across both moderate- and high-dimensional settings, along with varying levels of sparsity and noise. We then compared the difference in performance for each method across all scenarios between the simple and complex simulated datasets. Complete results for each scenario are provided in Supplementary Figures 3 and 4.

In terms of feature selection, all methods except random lasso performed better on the simple simulated datasets (Figure 5). BERM showed the smallest differences in performance between simple and complex datasets across both moderate- and high-dimensional datasets. In contrast, RENBOOT exhibited the largest performance difference in the moderate-dimensional setting, where it was the most accurate method for the simple simulated datasets (Figure 5 and Supplementary Figure 3). In moderate dimensions, the adaptive methods had variable performance: adaptive lasso was similarly accurate to BERM and RENBOOT on simple simulated datasets but experienced a sharp performance drop in complex scenarios. Adaptive elastic net, on the other hand, showed average accuracy on simple datasets with only a minor drop in complex datasets (Figure 5A and Supplementary Figure 3A). In high-dimensional settings, traditional and adaptive methods performed similarly on simple datasets but experienced declines in performance in complex scenarios. However, random lasso and RENBOOT showed only minor performance differences between simple and complex data due to their low performance in both settings (Figure 5B and Supplementary Figure 3B).

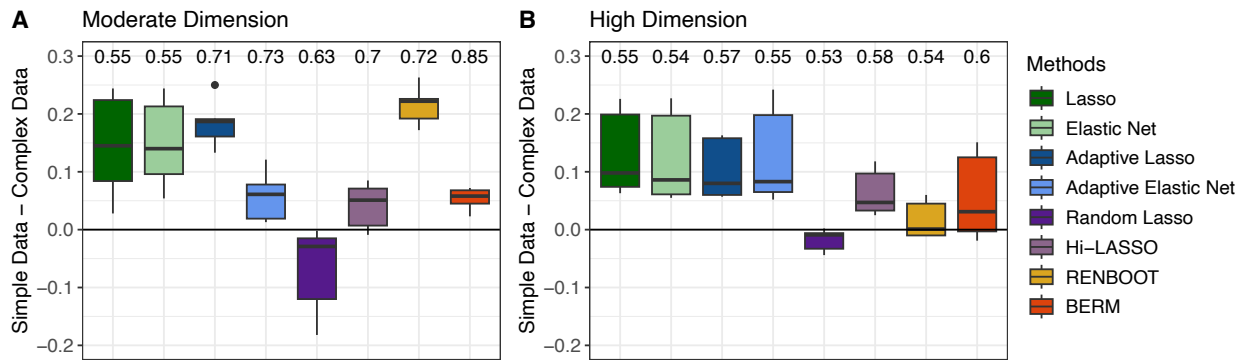

Figure 5. Difference in feature selection accuracy between simple and complex datasets across all scenarios. **A.** Mean difference in balanced accuracy between simple and complex datasets of correctly selecting non-zero variables (y-axis) for moderate-dimensional data simulated under various noise levels and model sparsities is shown for eight different methods. The mean balanced accuracy in complex datasets is shown at the top of the figure. **B.** Similar to A for the high-dimension scenario.

Regarding coefficient estimation on accurately selected variables, performance was generally better on the simple datasets (Figure 6). In the moderate-dimensional setting, all methods showed improvement, while in high-dimensional settings, all methods except lasso and adaptive elastic net demonstrated better performance. In the moderate dimensional setting, only Hi-LASSO and random lasso failed to accurately estimate coefficients in simple datasets and were increasingly poor in complex scenarios (Figure 6A). In the high-dimensional setting, the greatest performance differences occurred in the high-sparsity scenario, where the simple datasets had substantially lower MSE compared to those with lower sparsity levels (Figure 6B and Supplementary Figure 4B).

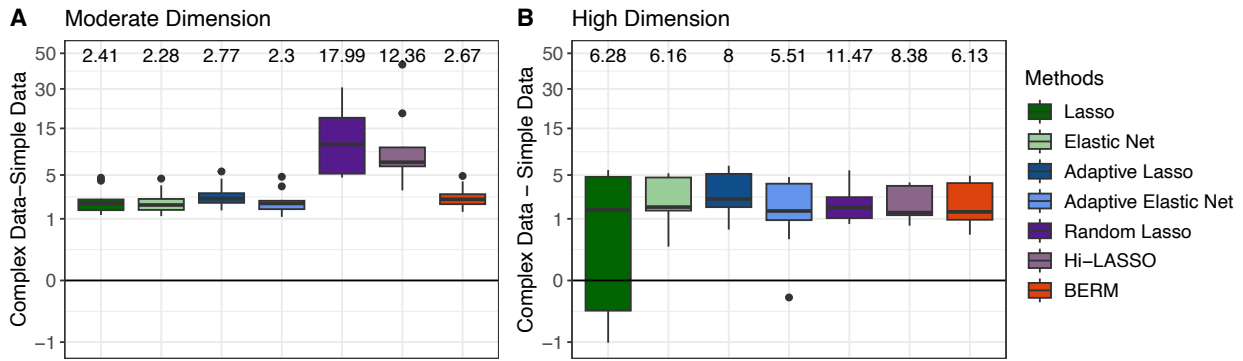

Figure 6. Comparison of mean coefficients estimation between simple and complex datasets across all sparsity and noise levels. **A.** Mean difference in MSE between simple and complex datasets for moderate-dimensional data simulated under various noise levels and model sparsities is shown for eight different methods. The mean MSE in complex datasets is shown at the top of the figure. **B.** Same as A, but for the high-dimensional scenario.

## 4. Case Study

### 4.1 Immunophenotyping Data

We analyzed previously published 192-parameter spectral flow cytometry data (Shapiro et al., 2023), which included samples from 252 peripheral blood mononuclear cell donors in a control group (CTR), defined as healthy individuals without diabetes who tested negative for Type 1 Diabetes-associated islet autoantibodies. The cohort also included 310 unaffected first-degree relatives of individuals with Type 1 Diabetes (REL), who may carry genetic or environmental predispositions related to the disease yet show no clinical signs of diabetes, and 240 participants diagnosed with Type 1 Diabetes (T1D). We applied BERM to the CTR group to identify features associated with immune aging with chronological age as the response and all features as predictors (see Supplementary Methods for additional details). We also fit the prediction model on both the T1D and REL groups, aiming to identify alterations caused by T1D and further investigate the impact of T1D on the immune system aging.

## 4.2 Results of Immune Age Model

We applied BERM to the control group of the immunophenotyping dataset, aiming to identify features associated with immune aging in the general population. BERM identified 84 features, achieving a test set prediction performance of $R^2 = 0.79$, compared to the original analysis (which utilized random lasso), that identified 69 features with a test set prediction performance of $R^2 = 0.70$ (Shapiro et al., 2023).

When using BERM to predict immune age in the T1D group, we observed that participants younger than 30 years with T1D exhibited significantly accelerated immune aging compared to the CTR and REL groups, with an average increase of 4.47 years (P < 0.001, Supplementary Figure 5). This represents a greater acceleration compared to our previously reported results, in which the average increase was 3.36 years (P < 0.001) (Shapiro et al., 2023).

There were 45 features selected by both approaches, and we further compared in terms of skewness, correlation relationships, and estimated coefficients for shared and uniquely identified features. BERM uniquely identified variables with extreme skewness (Figure 7A). Both methods were effective in identifying highly correlated features; some features uniquely identified by either BERM or random lasso exhibited high correlation with shared selected features (Figure 7B). However, features uniquely selected by random lasso were more likely to correlate with those selected by both methods (Figure 7C). All 45 features identified by both methods showed consistent directions in their coefficient estimates (Supplementary Figure 6). However, BERM was able to identify variables with subtle effects on immune aging, whereas random lasso, which employs a cutoff for feature selection, failed to recognize these lower-magnitude variables.

To further assess the relevance of the selected variables, we examined their association with immune age. In the CTR group, estimated coefficients for each immunophenotype from random lasso showed a positive correlation with their respective correlations to age, whereas BERM's coefficients did not exhibit as strong of a relationship (Figure 7D). This highlights key differences between the two approaches. Random lasso favored age-correlated predictors, as they consistently received larger coefficients across predictor subset bootstraps, making their final coefficients more reflective of univariate age associations. In contrast, BERM optimizes prediction by selecting variables and weighting them collectively in an elastic net. BERM's coefficients reflect feature importance within the model, and do not necessarily indicate univariate directional relationships with the outcome.

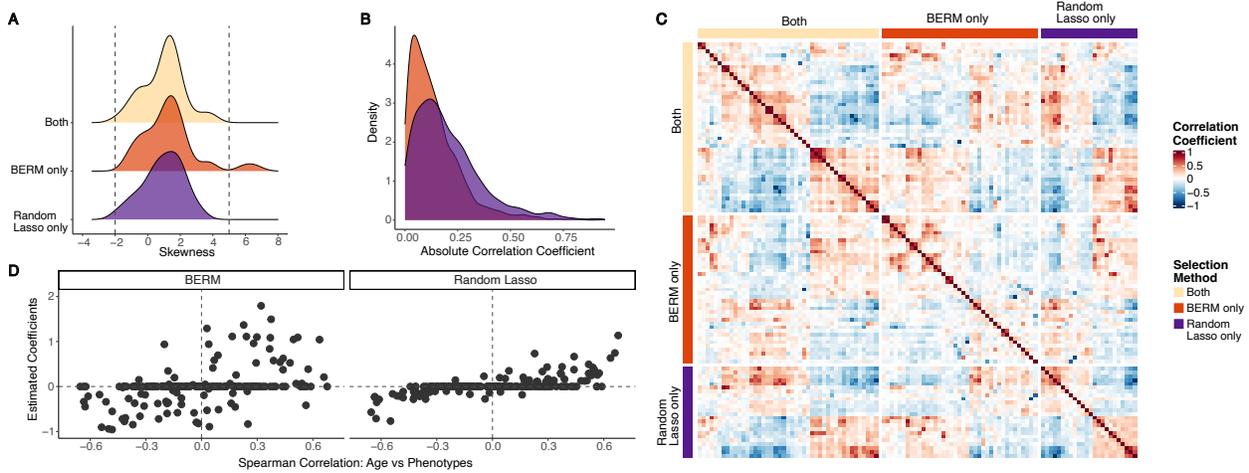

Figure 7. Comparison of feature selection and coefficient estimation by random lasso and BERM in the immunophenotyping dataset (Shapiro et al., 2023). A total of 45 immunophenotyping features were selected by both methods (noted as "Both"), while 39 features were identified exclusively by BERM (noted as "BERM only"), and 24 features were unique to random Lasso (noted as "Random Lasso only"). **A.** Skewness of selected features; **B.** Absolute value of Spearman correlation coefficients between features selected by both methods and those selected only by random lasso or BERM; **C.** Spearman correlation coefficients of selected features; **D.** Relationship between the estimated coefficients, as determined by BERM and random lasso, and their correlation with age.

**Discussion**

In the simulation studies, BERM accurately selected important variables and estimated their coefficients without clear overfitting. We observed that BERM had the most consistent performance across a wide range of scenarios in both moderate-dimensional and high-dimensional settings. Classical methods, including lasso and elastic net, performed poorly in feature selection for moderate-dimensional datasets, often selecting an excessive number of features—on average, 15 and 16 more, respectively. Additionally, classical methods and their adaptations, including lasso, elastic net, adaptive lasso, and adaptive elastic net, were heavily influenced by complex data structures. Among them, adaptive elastic net was particularly sensitive to noise and sparsity levels, suggesting a propensity to fit to noise rather than to the underlying signal. Random lasso, Hi-LASSO, and RENBOOT also had significant limitations, and appeared to perform relatively better in a limited range of simulation scenarios, specifically in highly sparse scenarios with low noise for the former two methods, or with high noise for the latter method.

BERM takes advantage of the superior performance of the elastic net in handling high multi-collinearity and improves upon robustness via bootstrap confidence intervals to significantly improve variable selection in complex data scenarios. On correctly selected variables, BERM's estimation of coefficients is comparable to that of lasso and elastic net approaches, allowing for accurate prediction. Although BERM incurs a higher computational burden compared to conventional methods like lasso and elastic net due to the use of cross-validation to tune optimal parameter $\lambda$, it is more efficient than alternatives such as random lasso and Hi-LASSO (Supplementary Table 1). We noted that BERM had reduced feature selection accuracy in extreme scenarios characterized by variables with either all zero coefficients or all non-zero coefficients (Supplementary Table 2). We found that tuning $\alpha$ via cross-validation rather than using the restricted approach of setting it directly at 0.5 in BERM's initial step, significantly enhanced feature selection in these challenging scenarios.

In the case study, BERM demonstrated superior predictive capabilities for estimating immune age. While random lasso provided biologically intuitive coefficients that reflected univariate age relationships, BERM's ensemble approach captured more complex, multivariable predictive patterns. This makes BERM particularly valuable in biomedical data analyses, where the goal is to obtain both a sparse model of important features along with an accurate prediction model. Additionally, given the uncertainty of the true sparsity model and noise-level in any data analysis, it is essential to have a method that delivers reliable performance across different scenarios. BERM outperformed all methods for feature selection across a wide variety of scenarios and demonstrated accuracy in coefficient estimation comparable to current leading approaches. These results highlight the potential of BERM as a robust tool for analyzing immunophenotyping data or similarly complex biomedical data.

**Data and Code Availability**

The simulated datasets are available on GitHub, along with all analysis codes for reproducibility (https://github.com/xiaorudong/berm-paper). The immunophenotyping dataset is available on Zenodo at https://doi.org/10.5281/zenodo.15189694.

**Acknowledgments**

The authors wish to thank Amanda Posgai from the University of Florida for her critical review of the manuscript. This work was supported by the National Institutes of Health (R35GM146895 to R.B. and P01AI042288 to T.B.).